\documentclass[a4paper,11pt]{article}
\usepackage{pos}
\usepackage[left]{lineno}
\usepackage{amsmath}
\usepackage{placeins}

\newcommand{\Nch}{\mbox{$N_{\mathrm{ch}}$~}}

\newcommand{\PbPb}{\mbox{Pb+Pb}}

\newcommand{\pp}{\mbox{$pp$}}
\newcommand{\pt}{\mbox{$p_{\mathrm{T}}$}}

\newcommand{\ptch}{\mbox{$p_{\mathrm{T}}^{\mathrm{ch}}$~}}
\newcommand{\ptz}{\mbox{$p_{\mathrm{T}}^{\mathrm{Z}}$~}}

\newcommand{\sqrtsnn}{\mbox{$\sqrt{s_{\mathrm {NN}}}$}}

\newcommand{\IAA}{\mbox{$I_{\mathrm{AA}}$}}

\newcommand{\kt}{\mbox{$k_{\mathrm{t}}$}}

\newcommand{\antikt}{\mbox{anti-\kt~}}

\newcommand{\xj}{\mbox{$x_{\mathrm{J}}$}}
\newcommand{\xhz}{\mbox{$x_{\mathrm{hZ}}$}}

\newcommand{\dndch}{\mbox{$(1/N_{\mathrm{Z}})(d^2\Nch/d\ptch d\Delta\phi)$}}

\title{New results on the modification of jet structure in Heavy Ion collisions}

\author*[a]{Helena Santos}
\author{on behalf of the ALICE, ATLAS, and CMS collaborations}
\affiliation[a]{Laborat\'orio de Instrumenta\c{c}\~ao e F\'isica Experimental de Part\'iculas, LIP,\\
 Av. Prof. Gama Pinto 2, Lisbon, Portugal}

\emailAdd{helena@lip.pt}

\abstract{These proceedings report recent studies performed by the ALICE, ATLAS and CMS experiments on jet substructure in \pp~and \PbPb~collisions at $\sqrtsnn$ = 5.02 TeV. The dependence of the modification of the jet shape on the dijet transverse momentum asymmetry is presented. Measurements of particles recoiling against the Z boson without being constrained to jet cone delimitations are shown. Fully-corrected measurements of the {\it N}-subjettiness within \antikt $R$ = 0.4 charged--particle jets are also described.}

\FullConference{%
 
 The Ninth Annual Conference on Large Hadron Collider Physics - LHCP2021\\
 7-12 June 2021\\
 Online
}

\begin{document}
\maketitle
Heavy ion collisions at ultra-relativistic energies take place at the Large Hadron Collider with the aim of studying the properties of the quark-gluon plasma (QGP)~\cite{paper:vanhove}. The hard scattered quarks and gluons emerging from collisions evolve as parton showers that propagate through the QGP. Constituents of these parton showers are expected to emit medium-induced gluon radiation. Consequently, the resulting jet loses energy -- a phenomenon commonly termed as $jet~quenching$~\cite{paper:jetquenching}. Accordingly, jets produced in heavy ion collisions are shown to be suppressed at a given transverse momentum, \pt, relatively to a sample produced in \pp~collisions~\cite{paper:raa}. Their internal structure is also shown to be modified~\cite{paper:FF}. Results shown in this conference used data produced in \pp~and \PbPb~collisions at the center of mass energy of 5.02 TeV, collected by three LHC experiments - ALICE, ATLAS and CMS. The Glauber Monte Carlo model~\cite{Glauber} is used to obtain a correspondence between the total transverse energy deposited in the forward detectores and the sampling fraction of the total inelastic \PbPb~cross-section, allowing the setting of the centrality percentiles.
\section{Jet shapes}
\begin{figure}[htb]
  \centerline{
    \includegraphics[width=0.3\textwidth, angle=0]{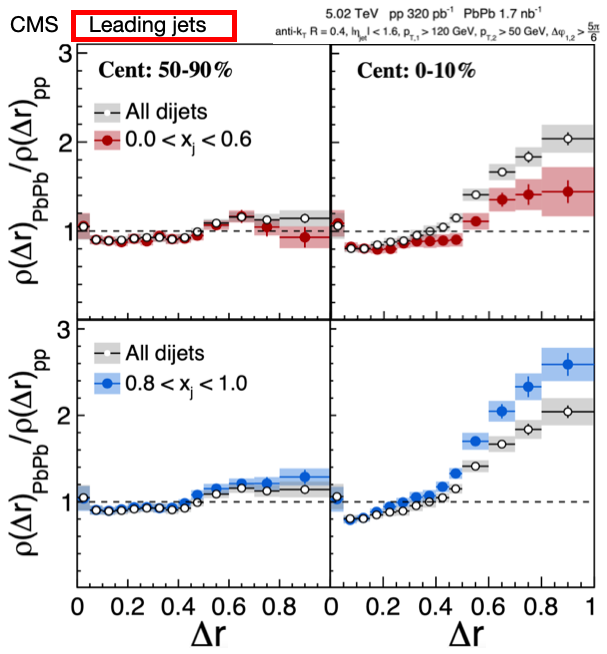}\\
    \includegraphics[width=0.3\textwidth, angle=0]{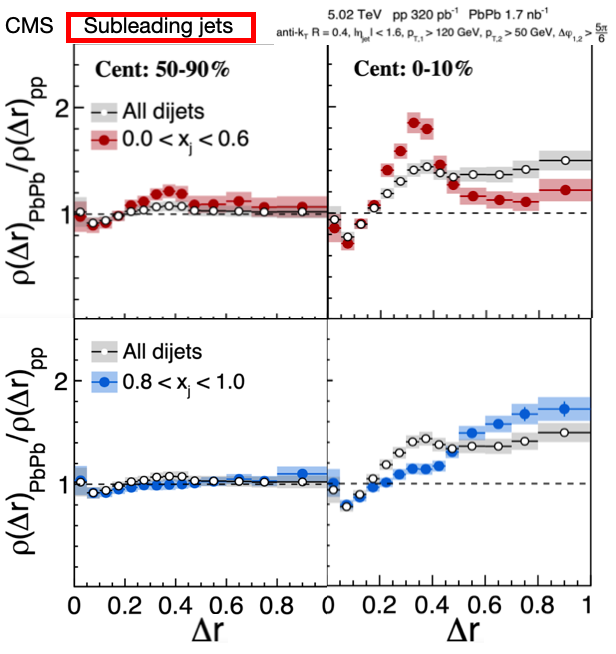}\\
  }
\caption{The \PbPb~to \pp~ratio as a function of $\Delta R$ for leading (left figure) and sub-leading (right figure) jet shapes, $\rho(\Delta R)_{\PbPb}/\rho(\Delta R)_(\pp)$, for $0 <\xj< 0.6$ (upper row) and $0.8 <\xj< 1$ (lower row) dijet selections. Peripheral (central) collisions are shown in left (right) panels of each figure respectively. The leading and sub-leading jet shape ratios for all dijets, i.e., without any selection on the dijet momentum balance, are also shown in each plot for comparison. The error bars represent the statistical uncertainties and the shaded areas the systematic uncertainties~\cite{paper:CMS_JetShapes}.}
\label{Fig:CMS_JetShapes}
\end{figure}

The CMS Collaboration analyses back-to-back dijets in \pp~and \PbPb~collisions via the ``jet shape'' observable, $\rho(\Delta R)$, which is the distribution of charged particles as a function of the angular distance from the leading and sub-leading jet axes, $\Delta R$. The goal is to understand how the modification of the jet shape relates to the degree of the asymmetry between the leading and sub-leading jet quantified by the ratio between the \pt~of the two leading jets, \xj~\cite{paper:CMS_JetShapes}.
Figure~\ref{Fig:CMS_JetShapes} shows the modifications of the leading and sub-leading jet shapes in \PbPb~collisions, assessed by the ratio of this observable in the two collision systems for two classes of dijets: asymmetric and symmetric. The black distributions represent all dijets without any selection on \xj. The modification of the leading jet shape for asymmetric dijets in peripheral \PbPb~collisions (50--90\%) is small. In central collisions (0--10\%), and for asymmetric dijets, the shape of the leading jet shows an increasing broadening with $\Delta R$ but remains less modified than the sample with all dijets. The opposite trend is observed in balanced dijets. The following interpretation is given: in asymmetric dijets the leading jet is likely to be produced close to the surface of the QGP, so emerging towards the detector without substantially modification. On the contrary, in case of symmetric dijets, both jets are expected to cross similar same path lengths, so losing approximately the same amount of energy. In this case the leading jet crosses the QGP and is strongly modified. Regarding the sub-leading jet in asymmetric dijets, a striking feature arises at the edge of the jet cone, $\Delta R$ = 0.4 - the protuberance already observed in peripheral collisions becomes exacerbate in central collisions. A possible interpretation is the existence of a third jet in \pp~collisions as required for momentum conservation, but which is absent in central \PbPb~collisions due to jet quenching. These results suggest significant effects from different paths length crossed and encourage phenomenological studies on medium response. 
\section{Modification of $Z$-tagged charged particle yields}
\begin{figure}[htb]
  \centerline{
    \includegraphics[width=0.3\textwidth, angle=0]{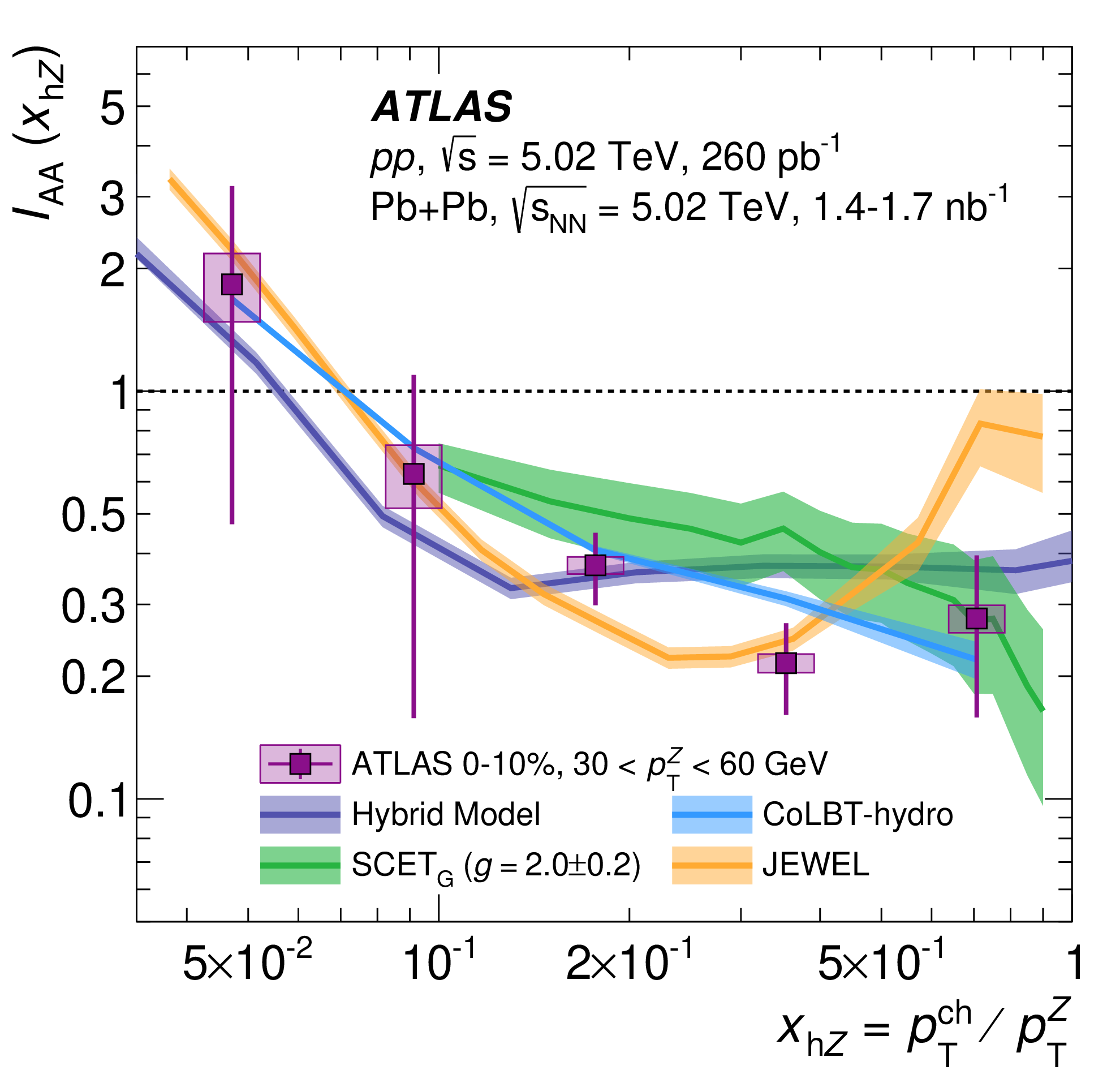}\\
    \includegraphics[width=0.3\textwidth, angle=0]{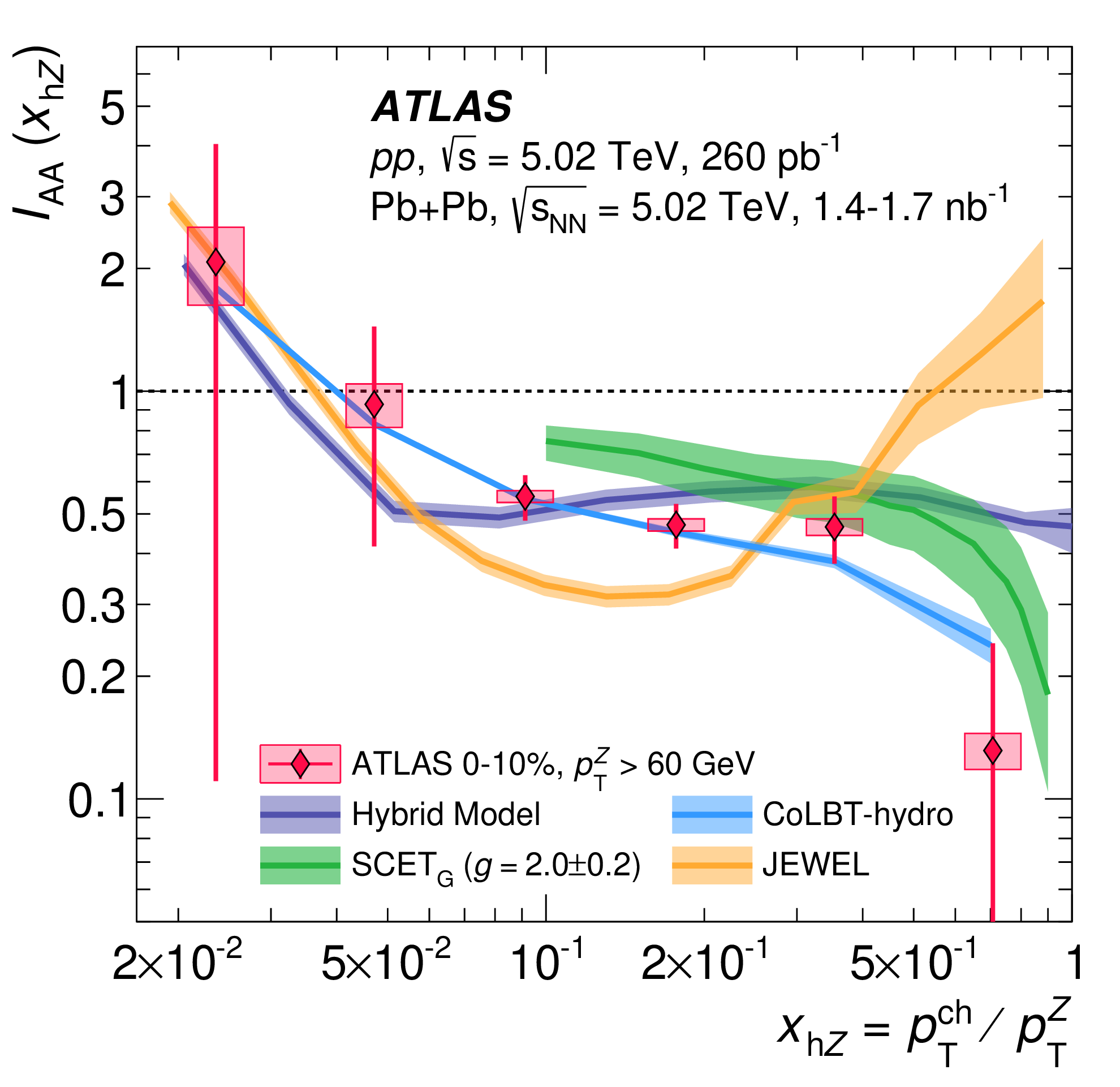}\\
  }
\caption{Ratio \IAA~of the per-$Z$ charged hadron yield in 0--10\% Pb+Pb events to those in \pp~events, for 30 $<$ \ptz $<$ 60 GeV (left) and \ptz $>$ 60 GeV (right). The data are compared with calculations in the Hybrid strong/weak coupling model (dark blue curves), JEWEL (orange curves), SCETG (green curves), and CoLBT-hydro (light blue curves). The vertical bars and boxes around the data correspond to the statistical and total systematic uncertainties. The shaded bands around the theoretical predictions represent the uncertainty associated with the theoretical predictions (statistical for JEWEL, Hybrid, and CoLBT-hydro, parametric for SCETG)~\cite{paper:Z, paper:Hybrid, paper:SCETg, paper:JEWEL2, paper:CoLBT-hydro2}.}
\label{Fig:xZh1}
\vspace{-0.4cm}
\end{figure}

The interest of studying jets recoiling against the $Z$ boson lies in the fact that the electroweak boson, not being affected by the QCD matter, is therefore an excellent tag for jet energy calibration, direction and flavour. However, the study of charged particles has the additional advantage of using events not constrained by the requirement of a reconstructed jet, which may result in a bias towards a sample with less energy loss than average.
The ATLAS experiment approach consists in counting the average number of charged--particles per $Z$ boson, \dndch, with \ptch$>$1 GeV and correlated with the $Z$ boson within an azimuthal opening angle $\Delta\phi>$3$\pi$/4. To quantify the modifications resulting from parton propagation through the QGP, the ratio \IAA~of the (per $Z$ boson) charged hadron yield in \PbPb~collisions to those in ~\pp~collisions is used. At fixed \pt, jets balancing $Z$ bosons arise from processes with different 4--momentum transfer squared Q$^2$. A test of the sensitivity of the energy loss process to parton virtuality is shown in Figure~\ref{Fig:xZh1}, where the ratio \IAA~as a function of the hadron-to-boson \pt~ratio, \xhz, is plotted for central collisions and for two ranges of \ptz. An enhancement at low values of \xhz~is observed followed by an increasing suppression with increasing values of this variable. One possible interpretation of these results is that partons produced with the $Z$ boson have initially higher \pt~values, but lose energy by interacting with the QGP medium, and therefore are re-distributed to lower values of \xhz. The mechanism of this energy loss is addressed by several models: the Hybrid jet quenching model~\cite{paper:Hybrid}; a model of jet charge modification in QCD matter (SCETG)~\cite{paper:SCETg}; the prediction generated by JEWEL (“jet evolution with energy loss”) model, which uses perturbative QCD to describe scattering of hard partons in a hot dense medium, coupled to a parton shower~\cite{paper:JEWEL2}; and the linear Boltzmann transport and hydrodynamics model (“CoLBT-hydro”), which simulates parton transport in a hydrodynamically evolving medium, including jet-induced medium excitations~\cite{paper:CoLBT-hydro2}. The results agree with the predictions, although Run 3 data will be fundamental to better constrain the models~\cite{paper:Z}.
\section{{\it N}-subjettiness}
\begin{figure}[htb]
  \centerline{
     \includegraphics[width=0.5\textwidth, angle=0]{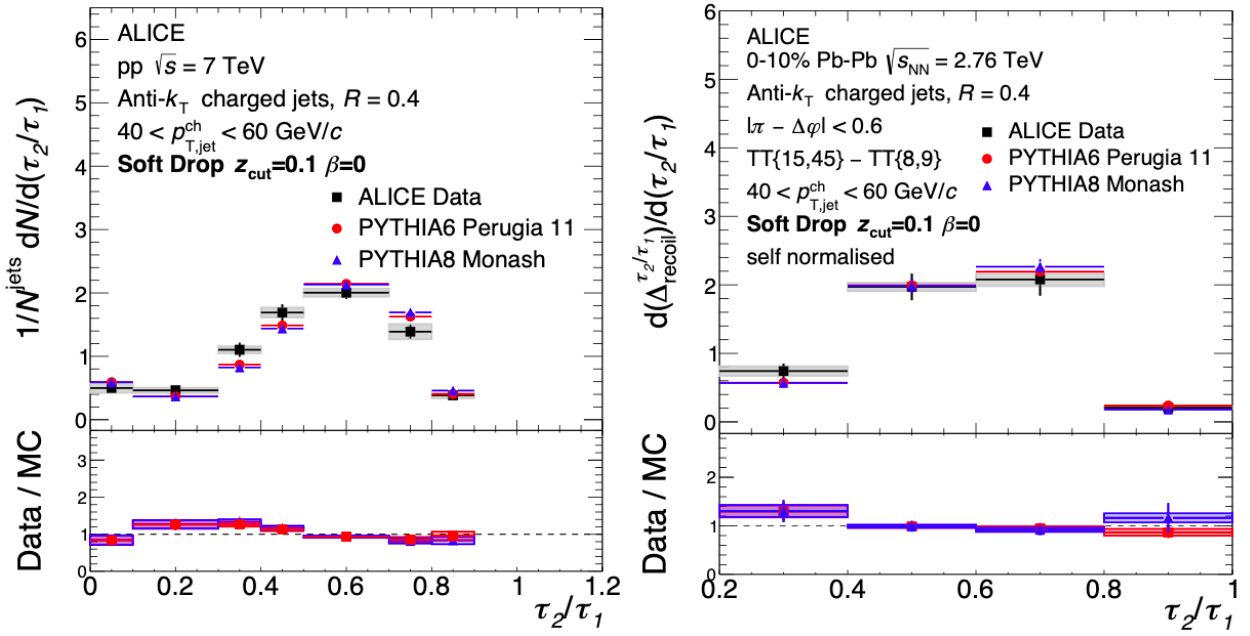}\\
  }
  \caption{Left panel: Fully corrected $\tau_2/\tau_2$ distributions, measured with C/A with Soft Drop grooming algorithm, in \pp~collisions at $\sqrt s$ = 7 TeV for \antikt $R$ = 0.4 charged--particle jets in the jet \ptch interval of 40–60 GeV/c. Right panel: same measurement but in \PbPb~collisions. The systematic uncertainties are given by the grey boxes. The results~\cite{paper:subjetiness} are compared with PYTHIA6 using the Perugia 2011 and Monash parameter tunes. The uncertainties presented for the PYTHIA distributions are purely statistical~\cite{paper:perugia}.}
\label{Fig:subjetiness}
\end{figure}

The ALICE experiment tests the ability of the QGP to distinguish between independent emitters via jet subjettiness within \antikt $R$ = 0.4 charged--particle jets. This observable is the \pt~weighted sum of the angular separation between each jet constituent and the closest subjet axis. The main idea is the reclustering of the jet seeking for two independent subjets and removal of non-perturbative constituents. The ratio of 2-subjettiness to 1-subjettiness, $\tau_2/\tau_2$, is sensitive to the rate of two-pronged jet substructure and it is shown in Figure~\ref{Fig:subjetiness}. The left panel shows the combinatorial suppressed jet distribution as a function of the $\tau_2/\tau_2$ in \pp~collisions where the Cambridge Aachen (C/A)~\cite{paper:C/A} algorithm with the Soft Drop~\cite{paper:SD} groomer applied has been used to recluster the jets. Within the kinematic range of the measurement (40 $<\pt<$ 60 GeV) the results favour a single-cored jet substructre. In fact both in \pp~and in central \PbPb~collisions the distributions peak at intermediate values suggesting that the two hard substructures cannot be well separated. The PYTHIA~\cite{paper:perugia} distributions appear slightly shifted towards less 2-prong jets compared to the data in \pp~collisions. The absence of this effect on the right panel hints at a reduction in the rate of two-pronged jets in \PbPb~central collisions compared to \pp~collisions. This analysis is comprehensively described in reference~\cite{paper:subjetiness}.\\

The author acknowledges the financial support of Funda\c{c}\~ao para a Ci\^encia e a Tecnologia (FCT) through FCT Researcher contracts CEECIND/03346/2017 and CERN/FIS-PAR/0002/2019.\\


\begin{thebibliography}{99}
  \bibitem{paper:vanhove}
L. Van Hove, CERN-TH-4204-85. DOI: 10.1016/0375-9474(86)90623-8
\bibitem{paper:jetquenching}
  J.-P. Blaizot, Y. Mehtar-Tani, \emph{J. Mod. Phys.} {\bf24}, 1530012 (2015).
\bibitem{paper:raa}
  ATLAS Collaboration, \emph{Phys. Lett. B} {\bf790} 108 (2019).
\bibitem{paper:FF}
   ATLAS Collaboration, \emph{Phys. Lett. B} {\bf739} 320 (2014).
\bibitem{Glauber} 
M.L. Miller, K. Reygers, S.J. Sanders, P. Steinberg, \emph{Ann. Rev. Nucl. Part. Sci.} {\bf57} 205 (2007).
\bibitem{paper:CMS_JetShapes}
  CMS Collaboration, \emph{JHEP} {\bf05} 116 (2021).
\bibitem{paper:Z}
ATLAS Collaboration, \emph{Phys. Rev. Lett.} {\bf126} 072301 (2021).
\bibitem{paper:Hybrid}
J. Casalderrey-Solana, D. C. Gulhan, J. G. Milhano, D. Pablos and K. Rajagopal, \emph{JHEP} {\bf03} 053 (2016).
\bibitem{paper:SCETg}
Y.-T. Chien, A. Emerman, Z.-B. Kang, G. Ovanesyan and I. Vitev, \emph{Phys. Rev. D} {\bf93} 074030 (2016).
\bibitem{paper:JEWEL2}
R. K. Elayavalli and K. C. Zapp, \emph{Eur. Phys. J. C} {\bf76} 695 (2016).
\bibitem{paper:CoLBT-hydro2}
  W. Chen, S. Cao, T. Luo, L.-G. Pang and X.-N. Wang, \emph{Phys. Lett. B} {\bf810} 135783 (2020).
\bibitem{paper:C/A}
  Y. L. Dokshitzer, G. D. Leder, S. Moretti, and B. R. Webber, \emph{JHEP} {\bf08} 001 (1997).
\bibitem{paper:SD}
  A. J. Larkoski, S. Marzani, G. Soyez, and J. Thaler, \emph{JHEP} {\bf05} 146 (2014).
\bibitem{paper:subjetiness}
ALICE Collaboration, arXiv:2105.04936 [nucl-ex].
\bibitem{paper:perugia}
  P. Z. Skands, \emph{Phys. Rev. D} {\bf82} 074018 (2010).
\end{thebibliography}
\end{document}